\documentclass{article}

\usepackage{arxiv}

\usepackage[utf8]{inputenc} 
\usepackage[T1]{fontenc}    
\usepackage{hyperref}       
\usepackage{url}            
\usepackage{booktabs}       
\usepackage{amsfonts}       
\usepackage{nicefrac}       
\usepackage{microtype}      
\usepackage{lipsum}
\usepackage{fancyhdr}       
\usepackage{graphicx}       
\usepackage{cite}

\title{Conceptual Design and Implementation of FIDO2 compatible Smart Card for Decentralized Financial Transaction System

}

\author{
Anisha Ghosh\\
Centre of Excellence, Artificial Intelligence \& Robotics (AIR),\\
School of Computer Science and Engineering\\
VIT-AP University, India \\
\texttt{ghoshanisha2002@gmail.com}\\
\And
Aditya Mitra \\
Centre of Excellence, Artificial Intelligence \& Robotics (AIR),\\
 School of Computer Science and Engineering\\
 VIT-AP University, India \\
\texttt{adityamitra5102@gmail.com}\\
\And
Sibi Chakkaravarthy Sethuraman \\
Centre of Excellence, Artificial Intelligence \& Robotics (AIR),\\
School of Computer Science and Engineering\\
VIT-AP University, India \\
\texttt{sb.sibi@gmail.com} \\
\And
Aswani Kumar Cherukuri\\
School of Computer Science and Engineering\\
VIT University, India \\
\texttt{aswani@vit.ac.in} \\
}

\begin{document}
\maketitle

\begin{abstract}
With challenges and limitations associated with security in the fintech industry, the rise to the need for data protection increases. However, the current existing passwordless and password-based peer to peer transactions in online banking systems are vulnerable to advanced forms of digital attacks. The influx of modern data protection methods keeps better records of the transactions, but it still does not address the issue of authentication and account takeovers during transactions. To the address the mentioned issue, this paper proposes a novel and robust peer to peer transaction system which employs best cloud security practices, proper use of cryptography and trusted computing to mitigate common vulnerabilities. We will be implementing FIDO2 compatible Smart Card to securely authenticate the user using physical smart cards and store the records in the cloud which enables access control by allowing access only when an access is requested. The standard incorporates multiple layers of security on cloud computing models to ensure secrecy of the said data. Services of the standard adhere to regulations provides by the government and assures privacy to the information of the payee or the end-user. The whole system has been implemented in the Internet of Things scenario.
\end{abstract}

\section{Introduction}
 Personal data such as banking details and passwords have been vulnerable to fall prey on the hands of numerous attackers and makes up for most reported cybercrimes in recent times. Many attempts have been made to protect sensitive data using passwords, pins and biometrics. However, all of them have been proven to be vulnerable to attacks. Most secure procedures have one or other vulnerabilities which can be exploited for unauthorized access to sensitive data and susceptible to loss of information \cite{malan2008managing,burnes2017prevalence}. \par
 Passwordless authentication implemented in this paper has been instrumental in overcoming the majority of attacks that traditional online securing methods such as passwords, pins and biometrics are susceptible to and are a threat to confidentiality. To enable transactions which assure security to end-user, we employ multiple security practices and keep all sensitive data encrypted. As a result, there should be the adoption of a stronger and secure authentication approach \cite{zwane2021intelligent}. The multiple security practices include enabling FIDO2 via smartcards across all supported browsers and platforms on billions of devices where FIDO2 cryptographic login credentials are unique across every website and biometrics or other secrets like cryptographic keys never leave the user’s device and are never stored on a server \cite{barnes2018crypto, anderson2021whom}. The proposed system incorporates advanced security features to ensure secrecy of the said data. With FIDO2 physical security keys, every user can have their own key or smartcard and it is easier to reset in case of loss of one. \par
To ensure complete security and prevent all the above vulnerabilities, we are keeping all data in transit encrypted using symmetric encryption keys which are instrumental in preventing man-in-the- middle and similar attacks. The private key never leaving the system ensures safety against most forms of digital attacks. Hence, in this paper the use of FIDO2 \cite{Nagasubramanian2020}\cite{vijayalakshmi2021impacts} bringing in the latest trends in cyber security for protecting sensitive encrypted data. \par
 PP2PP also makes use of strong identity verification (Identity Assurance Level 2) which will verify that the user is genuine in case someone else tries to appear as to be the user.  This security model eliminates the risks of phishing, all forms of password theft, cookie hijack, and replay attacks making it impossible to remotely enable or disable security keys. Security keys can be disabled when not needed so as prevent unnecessary use of it. Moreover, a single key can be used to manage all credentials of a particular person: a single FIDO2 compatible key can be used to authenticate to multiple online accounts and services. The cryptographic keys are stored on the hardware itself and never leave the trusted platform which ensures that an attacker cannot remotely access the account without physically possessing the card. The platform would be deployed on Cloud VMs and hence can be scaled up and down according to server load \cite{Yadav2019}. This is inclusive and accessible to all users regarding gender, ethnicity, disability, or other diverse characteristics if the user has access to a smartphone and an internet connectivity. \\
The contributions in this paper:
\begin{itemize}
    \item The paper proposes a novel and robust design standard for e-banking system that is ensuring complete security from malicious attacks using FIDO2 along with the implementation of decentralized blockchain architecture.
\item The proposed standard makes sure personal data does not fall in the hands of an adversary by making sure the requests are verified to be legitimate at every transaction.
\item The standard emphasizes passwordless authentication and multi-level seamless verification for registering an easy-to-use and completely secure banking system.
\end{itemize}
The rest of the paper is as follows: Section \ref{topic1:l1} discusses the background and related work of secure Online Banking systems. Section \ref{topic1:l2}presents the proposed standard for FIDO compatible smart payments card. Section \ref{topic1:l3} discusses the design implementation. Finally, section \ref{topic1:l4} is concluded our proposed work .

\section{Background and Related Work} \label{topic1:l1}
Most digital locks have leveraged PIN/Password security and/or biometrics for safekeeping of sensitive data. The security of e-banking is of topmost concern due to fraudulent behavior of attackers\cite{dhoot2020security}; the absence of strong e-banking security has kept numerous users skeptical of using e-banking services to avoid thefts such as removing money from a bank account and/or transfer money to an account in a different bank \cite{laborde2020user}. \par

Existing studies such as r using graphical passwords authentication techniques like recall and recognition has been made \cite{razvi2017implementation}, another paper uses graphical confirmation framework named pass grid to oppose bear surfing assaults with a one - time legitimate login marker and circulative even and vertical bars covering the whole extent of pass-pictures \cite{sukanya2017image}.It is to be noted that such existing banking systems have numerous vulnerabilities. For example, passwords and pins  once typed can be extracted by an attacker using suitable technology. One common way is the use of Forward Looking Infra-Red (FLIR) devices which are basically thermal cameras and can be used to capture the thermal residue of the user on the keyboard after typing. Similarly on touchscreen-based devices to enter passwords, pins, patterns, it usually leaves a smudge on the screen. This smudge can be captured even using a standard high-end mobile camera with proper settings. It has also been seen that it is quite difficult to remove the smudges, even after wiping  Another study proposed biometric online banking system as it tends to assist in lessening the cybercrime rate of online banking and tends to escalate the user confidence in using banking services online \cite{kiyani2020secure}. However, fingerprints authentication systems can particularly be vulnerable to a wide variety of attacks including using an artificial clone of the fingerprint, printed images, or even extracting the fingerprint from the authorized user by social engineering. Sometimes even restricting access to a particular person may imply changing the lock pin or password for everyone implying such methods are highly inconvenient.\par

 Existing studies in the literature shows that FIDO UAF and World Wide Web Consortium (W3C) Verifiable Credentials can be used to present a user-centric and decentralized digital identity system \cite{zwane2021intelligent,razvi2017implementation}. It has made digital identity highly trustworthy both for the user and the service provider who may be authenticating the user. The entire system was implemented for a banking scenario to show how secure it could be, and has also allowed users to generate on-demand identities that could contain only the necessary information \cite{zwane2021intelligent}. Their model presented the service provider with the authenticated information from the source directly. Another paper presented the application of FIDO protocol to enable multi-factor authentication in banking scenarios. It allowed a single gesture phishing-resistant multi-factor authentication. It involves the keys and biometrics to stay on the user’s device and no server-side secrets. It also ensures no thirdparty protocol is involved \cite{sukanya2017image}. A study proposed a promising approach to maintain security even after a FIDO authentication is done. A continuous FIDO authentication browser extension allows the Relying Party (RP) and the authenticator to continuously exchange verification in the background. Hence in this paper we propose PP2PP, the first to introduce peer to peer secure transactions using FIDO2 physical security. Table \ref{tab:my-table1} compares PP2PP with other conventional methods followed by payment industry.\par
 
\begin{table}[]
\centering
\caption{Comparison of PP2PP vs other payment systems }
\label{tab:my-table1}
\resizebox{\columnwidth}{!}{%
\begin{tabular}{llll}
\hline
\multicolumn{1}{c}{\textbf{Research}} &
  \multicolumn{1}{c}{\textbf{Conventional     Methods}} &
  \multicolumn{1}{c}{\textbf{TIQR}} &
  \multicolumn{1}{c}{\textbf{Current paper  – PP2PP}} \\ \hline
Security Logic          & Password/Identity    & PINs, OTPs           & Passwordless                                                                      \\ 
Authentication protocol & Application Specific & QRs                  & \begin{tabular}[c]{@{}l@{}}Web Authentication \\ (WebAuthn) and CTAP\end{tabular} \\ 
Communication protocol  & HTTP                 & HTTP/HTTPS           & HTTPS only                                                                        \\ 
Library utilized        & Application specific & Application specific & \begin{tabular}[c]{@{}l@{}}FIDO2, AES symmetric \\ Cryptosystems\end{tabular}     \\ \hline
\end{tabular}%
}
\end{table}
In this paper we are integrating multi-factor authentication with Fast Identity Online (FIDO) so that PP2PP  does not have any of the vulnerabilities discussed above by enabling passwordless authentication using physical security and device attestation in a client-server model \cite{zwane2021intelligent}. In terms of digital asset security, FIDO2 is a new age technology for MFA and Passwordless authentications. FIDO2 leverages Client to Authenticator Protocol (CTAP) \cite{roy2017smart} and public key cryptosystem for secure authentication. FIDO is a relatively newer technology \cite{zwane2021intelligent} and there is a high possibility that it will be a highly used authentication technology online in near future. This study can be easily executed without any hassle of complex pin/password settings and configurations. The proposed standard is more seamless as it implements device attestation for authentication, which is a passwordless model \cite{klieme2020fidonuous,lyastani2020fido2}. Device attestation changes the authentication model to ‘something the user has’ instead of ‘something the user knows.’\cite{Nagasubramanian2020}.  It is proven to be secure against most common attacks and care has been taken to patch common vulnerabilities.\par 
\begin{table}[]
\centering
\caption{Comparison of PP2PP with the recent relate works}
\label{tab:my-table2}
\resizebox{\columnwidth}{!}{%
\begin{tabular}{lccccc}
\hline
\multicolumn{1}{c}{\textbf{Features}} & \textbf{PP2PP} & \textbf{CASH} & \textbf{CARDS} & \textbf{BANK-TRANSFER} & \textbf{MOBILE-PAYMENTS} \\ \hline
Fast to use                                                                         & YES & YES & NO         & NO         & YES        \\ 
Easy to register                                                                    & yes & na  & no         & No         & no         \\ 
\begin{tabular}[c]{@{}l@{}}Support for biometric  \\  authentication\end{tabular}   & yes & NA  & no         & No         & yes        \\ 
Secure against remote   hijack                                                      & yes & yes & no         & No         & no         \\ 
Secure when card is stolen                                                          & yes & na  & vulnerable & vulnerable & vulnerable \\ 
\begin{tabular}[c]{@{}l@{}}Same key for protecting   \\ digital assets\end{tabular} & yes & na  & na         & no         & no         \\ \hline
\end{tabular}%
}
\end{table}
 To enable transactions using this standard , the physical security key is needed. During registration, the keys are enrolled by each user. To further add biometric authentication, a fingerprint compatible physical security key or a smart card communicates with the client device through Near Field Contact (NFC), Bluetooth Low Energy (BLE) or USB. It has made digital identity highly trustworthy both for the user as well as the service provider who may be authenticating the user because the card. This is because it allows a single gesture phishing-resistant multi-factor authentication and involves the keys and biometrics to stay on the user’s device itself and has no server-side secrets. Even if the smart card is stolen, the private keys or fingerprint matches will prevent all malicious attacks. Another research presented a large-scale lab study of FIDO2 single-factor authentication and collected insights about the perception, acceptance, and concerns about passwordless authentication among the end-users. Their results showed that users are willing to accept a replacement of text-based passwords with a security key for single-factor authentication \cite{van2011tiqr}. Table \ref{tab:my-table2} shows a comparison between PP2PP and other payment methods such as cash, cards, bank transfer and others\par 
 
 Tiqr \cite{van2011tiqr} is an authentication framework based on a smartphone app which also implements multifactor authentication. The application scans a QR code displayed on websites which contains a challenge to be answered by the smartphone after entering the correct PIN. However, the challenge is an OTP sent to the website by the smartphone. Therefore, the user does not need to handle the code manually. However, this does not differ substantially from use of passwords, pins, face IDs , etc. and might be susceptible to attacks as well. Hence, in our idea we can claim FIDO2 which only makes use of a smart card implementing physical security keys inbuilt with fingerprint sensors and U2F authentication . The secondary way, if the first is not successful, is using verification link sent to the email of the user. The use of symmetric cryptosystems to verify cookies storing old and new information will assure no information is being phished. The data stored by cookies stays on the system and hence cannot be susceptible to server side attacks and can neither be hijacked successful by stealing information from the system because each time the user tries to make a transaction, cookies will be verified.
 
 \section{Proposed Solution and Novelty of PP2PP} 
 \label{topic1:l2}
 
It is evident from above discussion that PP2PP is successful in eradicating all possible vulnerabilities and can be claimed as the most secure and revolutionary banking management system. Instead of going through multiple steps of verification with time consuming and vulnerable systems like pins and passwords we would be implementing FIDO 2 specifications which uses a single step seamless authentication.\par
Since the use of passwords have been proven to be vulnerable to several attacks as discussed above, we are proposing a  prevalent way of implementing multi-factor authentication. One of the ways of implementing this  is by using one time passwords (OTPs). Using OTPs as strong second factors is generally based on possession of a device, e.g. a particular hardware token or an application on a mobile phone. Generation of the OTPs may rely on time synchronization between client and server or on a secret seed and an algorithm to generate a chain of passwords (HMAC based OTPs and time-based OTPs, respectively). Banking applications generally refer to OTPs as transaction authentication numbers (TANs). Theses TAN are often forwarded to the user through a third channel, for example via SMS. Other approaches use special hardware tokens to generate OTPs. However, OTP systems are vulnerable to attacks as well. Thus, implementing multi factor authentication using public and private keys are most secure as even if the smart card is stolen, the private key never leaves the device.\par
There are some standards implementing Universal Second Factor and hardware-backed tokens for authentication. First, we enable the use of existing easy to use smart card along with providing increased token mobility by enabling remote authenticator tokens. The physical security key or smart card signs it with the private key after verifying the authenticity of the relying party which ensures no third party has access to the user’s details and establishes the fact that even if the physical key is stolen, the private key never leaves the system and is impossible to fall prey into the hands of attackers. Our showcases demonstrate logging into a web application of PP2PP using  a FIDO2 smart card . 

\subsection{Proposed Work}
We created several showcases to demonstrate the capabilities of our approach. In the proposed standard, security and privacy is achieved through employment of multiple levels of secure practices and isolating the components possess part of the knowledge to derive secrets from system or web server. PP2PP used the smart card which can be used to authenticate the user to the portal using CTAP protocol (Client to Authenticator) protocol which includes use of Bluetooth, NFC or USB as it  defines how to establish communications between FIDO2-enabled browsers and operating systems.\par
PP2PP offers a methodology for using physical smart card to protect digital information. Robust cloud security principles and in transit data security is provided against most known attacks. Moreover, incorporating a public key cryptosystem and trusted computing ensures keys cannot be cloned. In every transaction request new tokens are generated in combination to eliminate attacks. From a legitimate user’s point of view , using PP2PP is as simple as using a card to enable transaction. The difference is that in case of PP2PP multiple levels of security checks take place in a very fast and seamless manner.

\subsection{Design Overview of PP2PP}

\begin{figure}
  \centering
  \includegraphics[width=1.05\textwidth]{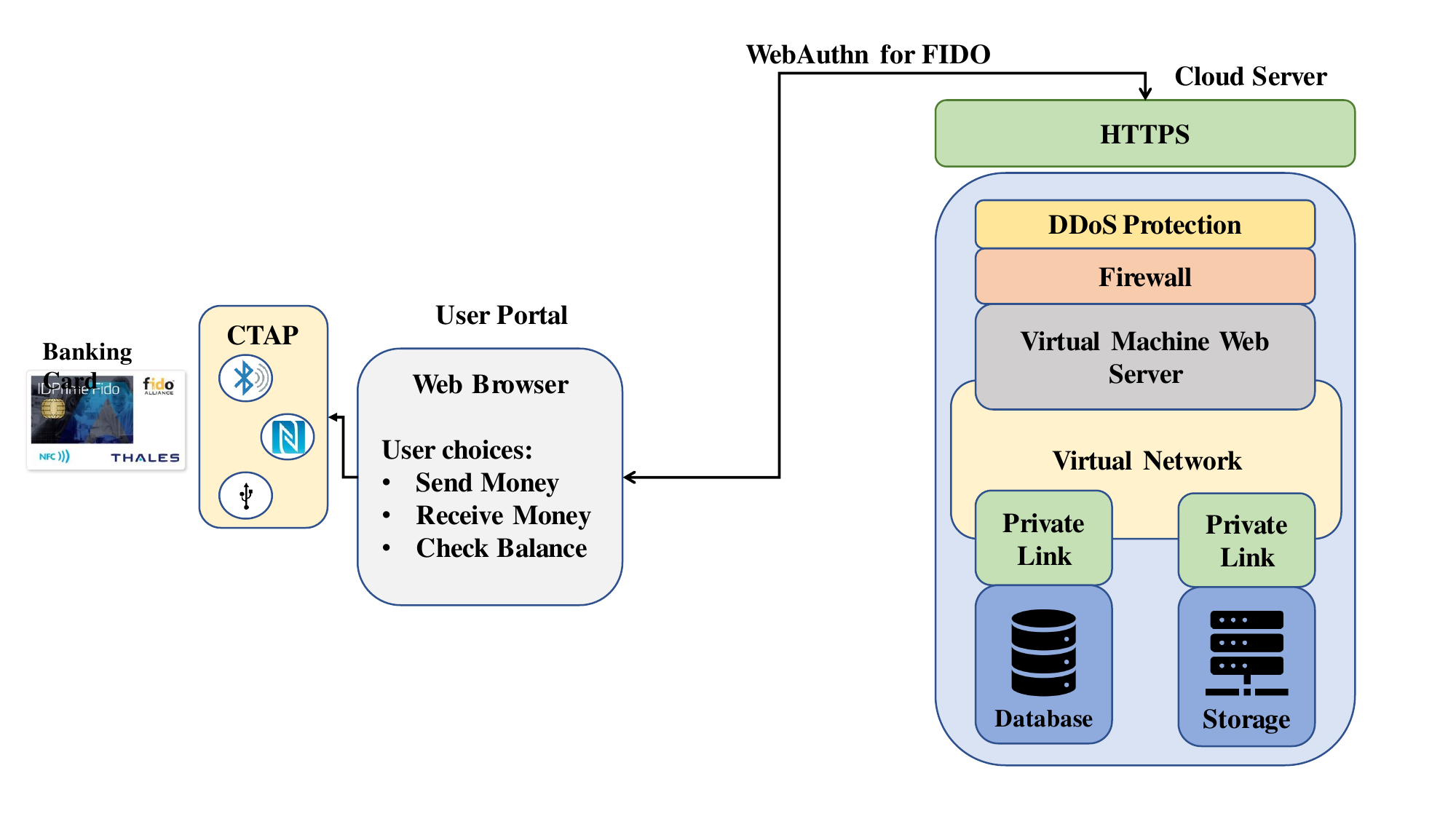}
  \caption{Architectural overview of the proposed framework.}
  \label{fig:fig1}
\end{figure}

Fig. \ref{fig:fig1} shows an architecture overview of the proposed framework. The first layer starts with the user, and subsequently the client through which all communication is made to the servers.

The proposed system essentially works on the Cloud and Edge computing models. Using steps mentioned above, the user interacts with the web browser, preferably Mozilla Firefox which runs in the background and the web browser interacts with the cloud architecture. The cloud architecture uses the following:\\

\textbf{HTTPS:}The first layer of the server uses  HTTPS protocol which transfers data by establishing a secure connection. All communication between the client and servers goes through HTTPS.\\
\textbf{DDoS:}DDoS protection drops all packets with is suspected to be malicious and aimed at flooding the system.\\
\textbf{Firewall:} The firewall on the other hand allows only traffic encrypted by SSL/TLS on port 443 and doesn’t allow any non encrypted packet to interfere. \\
\textbf{Virtual Machine Web Server:} This is the Apache web server which is the most popular and most used HTTP server platform responsible to deliver web content through the internet\\
\textbf{Virtual network:} The Then virtual network is a separate subnet containing the Virtual Machine, SQL server and blob storage. \\ 
The virtual machine connects to SQL server(database) and blob (storage) through private links which are similar to VPNs so that no outsider can intercept any information.\par
To implement the proposed specification, following parties are involved:
\begin{enumerate}
   \item \textbf{Sender:} The sender refers to the party whose information is to be stored safely and securely whilst protecting their privacy. This is the user who has the knowledge of the passphrase and can provide credentials unique to them. The sender can perform following actions.\\
 1.1	Check transaction history\\
1.2	Create a sharing token which can be shared over NFC Cards, nearby share and so on\\
 1.3	To send money directly\\
\item \textbf{Bank:} Bank refers to the service responsible for storage, retrieval and presenting sensitive data to and from the user. The bank can choose to adhere to the proposed standard and initiate secure information exchange transactions with the trusted third party. The bank can perform the following actions\\
1.1	Authenticate a new user\\
 1.2	Register for a smart card\\
 1.3	Replace the smart card enabled with FIDO2 specifications if lost\\
 \item  \textbf{Receiver:}The receiver refers to the other party on the other end of the application who requests for payment and whose information is also to be stored safely and securely whilst protecting their privacy
\end{enumerate}
\subsubsection{Primitives}
The following primitives are the key components of the system:
\begin{enumerate}
\item   \textbf{Physical Security Key:} This is a physical device or the trusted computing platform of the user’s device that can store the private key of public-key cryptosystem securely on hardware and can communicate with the web browser using the Client to Authenticator Protocol (CTAP).
\item     \textbf{Key:} This is the cryptographic key used to encrypt sensitive information. RSA, AES128 and AES256 cryptosystems have been used in this study effectively.
\item    \textbf{Relying Party/Cloud server:} This is the web server running on the cloud virtual machine. This is used as a relying party for FIDO2 specifications and a web server for the other operations. All traffic from the public internet to the Cloud server and vice versa is encrypted with SSL.
\end{enumerate}
\subsubsection *{Cryptography, Hashing and Keys}
Cryptography and Hashing are the key components of the study when it comes to providing security to the personal account.
\begin{enumerate}
\item  \textbf{Authentication keypair:} This is a keypair unique to each physical security key. It follows the RSA Cryptosystem. The private key is stored securely in the hardware of the physical security key. The public key is stored in the form of Binary Large Object (BLOB) in a storage account associated with the cloud server.
\item  \textbf{Application key:} This key is used to encrypt all sensitive data when being served on the webpage in the cookies. This follows AES128 cryptosystem, and this key is saved on the disk of the virtual machine.
\end{enumerate}

\begin{figure}
  \centering
  \includegraphics[width=1.05\textwidth]{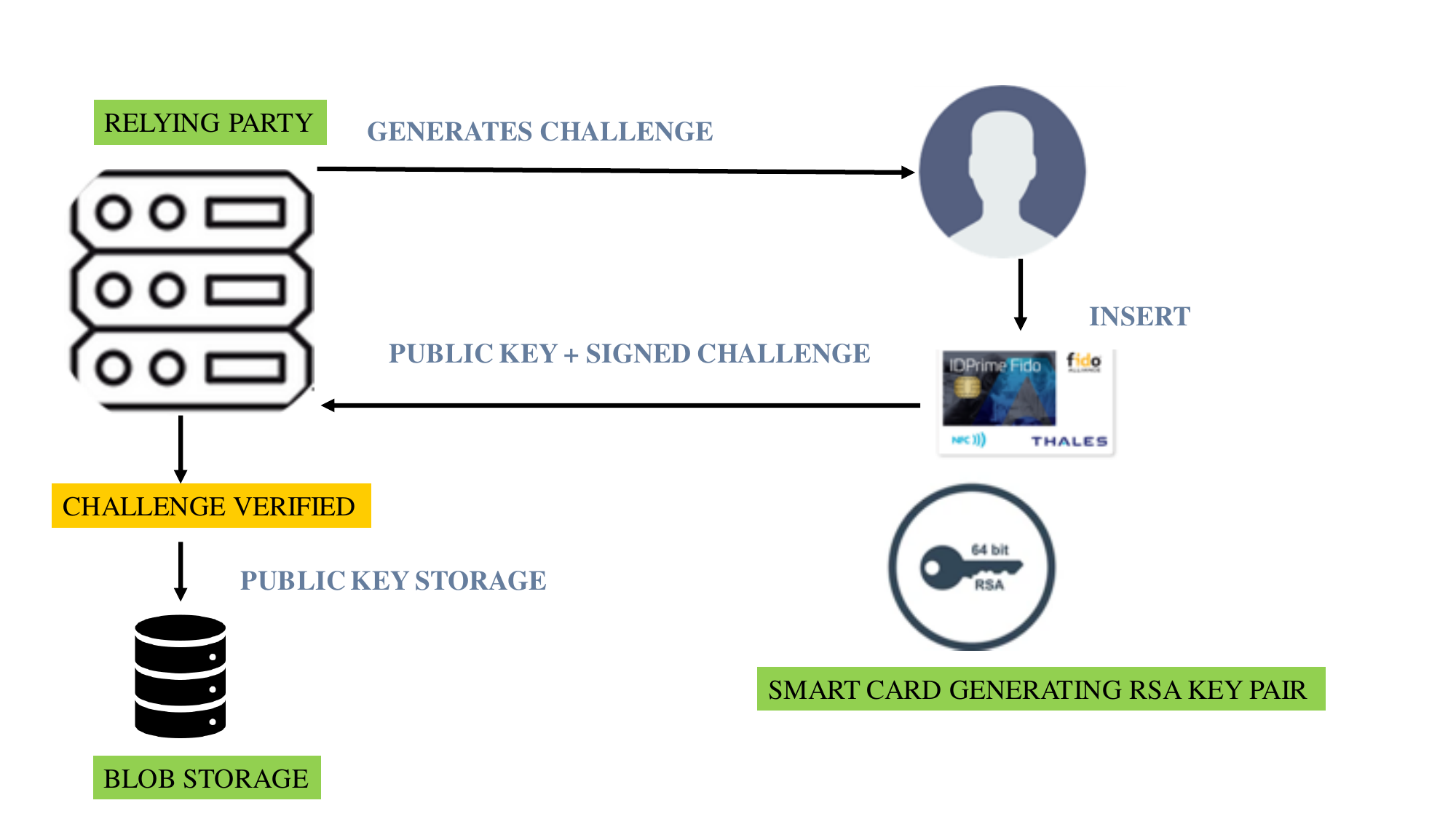}
  \caption{Registration workflow of the proposed system}
  \label{fig:fig2}
\end{figure}

\subsection{Device Registration}
For device registration, the Relying Party creates a challenge and sends it to the user. The user connects his smart card. An RSA keypair is generated in the smart card. The challenge is signed with the private key. Then the public key and the signed challenge is sent back to the Relying Party. The Relying Party verifies the signed challenge and saves the public key against the user in BLOB storage. The Relying Party verifies the signed challenge and saves the public key against the user in BLOB storage as shown in Fig. \ref{fig:fig2}. \par 

\textbf{Authentication:} Authenticating a user is done in two different ways. The primary way is passwordless authentication using FIDO2 compatible smart cards or physical security keys. The secondary way, if the first is not successful, is using verification link sent to the email of the user.\par

\textbf{Passwordless login with FIDO specifications:} FIDO2 specifications are based on W3C WebAuthn for securely communicating the Cryptographic keys and CTAP for communicating with the hardware module storing the private key. The RP server generates a challenge which is passed on to the client browser using WebAuthn. It is then passed on to the Physical Security key using CTAP. The physical security key communicates with the client device via NFC, BLE or USB. The physical security key signs it with the private key after verifying the authenticity of the RP. Then signed challenge is then returned and the RP verifies it using the saved public key. It is to be noted that the private key is stored securely in the hardware only. \par
\textbf{Passwordless login with one time links:} A one-time link is sent to the email of the user. This link contains an authentication token (UUID version 4) and the requester IP address encrypted together. The requested IP address is verified once the user opens the link, and the username is fetched from the database by querying it with the registration data.
\subsubsection {Transaction flow}
This is the main stage of any payment system, in which the transfer of  values takes place. It typically involves four consecutive sub steps as discussed below \cite{de1998scheme}.
\begin{enumerate}
\item 	\textbf{ Authentication:} It may be necessary for the parties of a transaction to identify themselves, to ensure that each is an authorized user of the system.
\item  \textbf{ Payment:} The user needs to choose to receive money or send money. If user chooses to  send money, they can do so in 4 ways: \\
 2.1. QR sharing \\
              2.2. Tokenizing with NFC tokens\\
              2.3. Link sharing, including nearby share.\\
2.4.	Direct to account. \\
\item \textbf{ Acknowledgment:} The payee or receiver needs to accept the payment request sent may send the payer a receipt corresponding to the transaction.
\item  \textbf{ Conclusion:} If all parties agree on the outcome of the transaction, it can be ended without problems. If there is any disagreement, a dispute resolution protocol may be started.
 Settlement. The amount is readjusted with the bank balance and displayed in dashboard.\\
 \item \textbf{ Desirable Requisites Integrity:} partially guaranteed. Digital signatures are used to maintain certificate integrity. Tokens are not protected by integrity mechanisms. 
\end{enumerate}

  \begin{figure}
  \centering
  \includegraphics[width=1\textwidth]{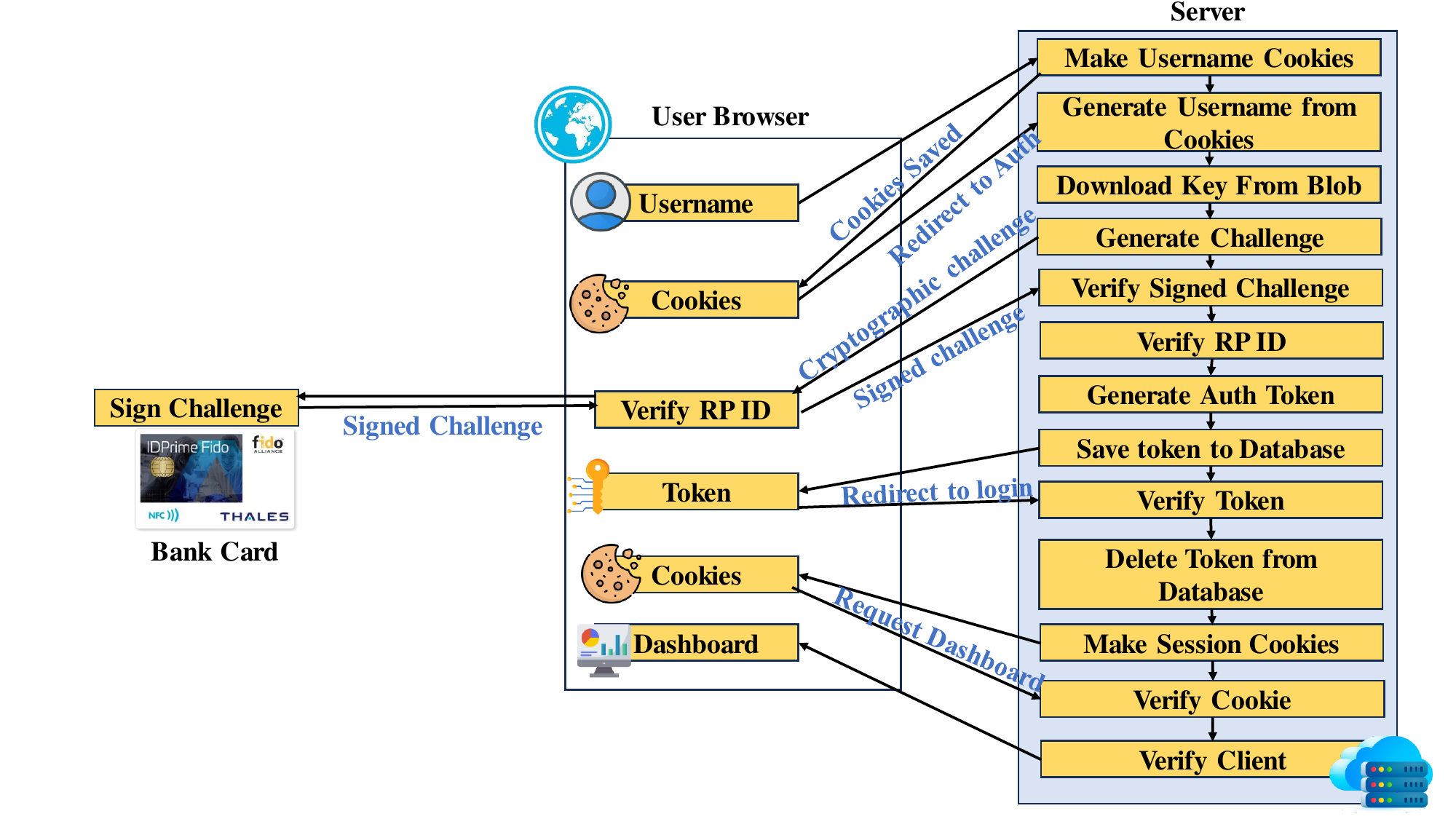}
  \caption{Authentication workflow.}
  \label{fig:fig3}
\end{figure}

\par
In Fig. \ref{fig:fig3}, the user authenticates himself with the FIDO 2 enable smart card embedded with NFCs, BLEs or USBs. Once they login with the card, a signed challenge is generated towards the user’s browser to verify his credentials or the relying party ID. The relying party sends the same signed challenge to the server and on receiving, the cloud server verifies the signed challenge with the original information stored. \par
As the username is entered, the information is shared with the server and cookies containing this information is saved in the local system. After cookies are saved, user is redirected to the authorization page to generate username from cookies for verification with the original that was saved in the local system and currently entered data as above.\par
After this, the key from azure blob storage is downloaded and challenge is generated to verify the relying party ID in the cloud server itself. After verification is done, an authentication token is generated in there and the token is saved in database. This token is then redirected to the web browser which consists details of the user according to the previously stored information or token. The browser token and server token is verified in the cloud server and  the token is deleted after verification is successful.
The server then creates session cookies and collects the data of cookies stored on user’s system or in the web browser with the originally stored data in cookies on the server side. Once verification is over, the client is verified from the dashboard and authorization is successful.\par

\section{Experimental Setup} 
\label{topic1:l3}

The cloud server is running on a standard Azure B1s Virtual Machine with 1GiB RAM and 1 vCPU. It is running Ubuntu Server 18.04-LTS. It has been secured with firewall and network security groups to keep malicious users out. The web server is running on Apache2 and is interfaced with a Web Server Gateway Interface (WSGI) to the program written using Flask library. Proper cloud security practices are followed to ensure desired transactions.

\subsubsection{Attack Analysis}       

 Malware: The virtual machine is scanned in real-time by cloud anti-malware service for any known threat or malware and if found, is mitigated.\par
 
\textbf{Phishing:} Phishing is not possible in this model as  FIDO2 specifications enforce Relying Party validation at every step. Moreover, A phished website cannot replay the FIDO key inputted by the legitimate user. SSL certificates are used to verify ownership of the backend server and a phished site requesting a FIDO key will result in an error on the client-side. For example, when a request is made to a phished server requesting the FIDO key with the RP of the legitimate server, it would result in an Invalid Domain error.\par

\textbf{Man-in-the-middle attacks (MITM):} MITM attacks are mitigated as all communications between the client and the server are encrypted with SSL. \par

\textbf{Distributed Denial-of-Service (DDoS):} The virtual machine is protected by DDoS protection. It is covered with active traffic monitoring and always-on detection. Moreover, automatic attack mitigation is helpful for backend resilience. This has been tested with the Hulk tool to increase the server’s load and monitor it with cloud insights. It is observed that after a high number of requests are made, the DDoS protection blocks the incoming requests from the client attempting to overwhelm the server while normal operations remain unaffected.\par

\textbf{SQL Injection:} Advanced threat protection for the SQL Database detects anomalous activities indicating unusual and potentially harmful attempts to access or exploit databases. Advanced Threat Protection can identify Potential SQL injection, access from unique locations or data centre, access from an unfamiliar principal or potentially harmful application, and brute for SQL credentials. Moreover, the firewall of the SQL database allows access only from the virtual machine. Thus, the databases are protected against all known forms of attacks.\par

\textbf{Cookie/Session hijack:} The cookies are appended with the client IP address and encrypted with AES256. Hence, the adversary even on hijacking the cookie, cannot modify or alter it. And the client IP address is verified from the cookie at every operation. Thus, cookie or session hijack is mitigated.

\subsection{Performance Analysis}

It is seen that PP2PP takes significantly less time, about 400 ms (plus user input times), for registering and authenticating the  security keys. Hence, it can be claimed that it is quite efficient and can be implemented in real-world scenarios. 5 keys have  been registered and authenticated. The time taken for the operations (Key registration , authorization and transfer). are given in the Table. \ref{tab:my-table3} and Table. \ref{tab:my-table4}. The analysis was done with the edge device which have an internet connectivity of about 3 Mbps. However, the challenge being of 16 bytes, it is tested to be working under low internet conditions. \par

The backend virtual machine is on elastic scaling that scales the backend VMs up or down when needed. It is further going through an elastic load balancer which is instrumental in distributing the load among all the VMs in the VM set. Availability sets and Availability zones make the backend resilient to most outages. According to the Service Level Agreements (SLA), the monthly uptime should be over 99.99\% and using Availability sets, Availability zones can be instrumental in keeping the service up even in case of a disaster or a catastrophic failure in a data center. The edge device in the lock will work as long as there is internet and power connectivity. For uninterrupted usage, proper error and exception handling and recovery have been implemented in both the edge devices and the cloud VMs. Thus, it is claimed that PP2PP, in the experimental setup, runs fast enough and is reliable enough against downtimes and failures.

\begin{table}[]
\centering
\caption{Key registration time}
\label{tab:my-table3}
\begin{tabular}{cccc}
\hline
\textbf{Sl.No.} &
  \textbf{\begin{tabular}[c]{@{}c@{}}Time to   create\\    challenge\end{tabular}} &
  \textbf{\begin{tabular}[c]{@{}c@{}}Time to   verify  \\ and register key\end{tabular}} &
  \textbf{\begin{tabular}[c]{@{}c@{}}Total   time    for   \\ processing\end{tabular}} \\ \hline
1       & 302ms   & 288ms & 590ms   \\ 
2       & 311ms   & 277ms & 588ms   \\ 
3       & 311ms   & 64ms  & 375ms   \\ 
4       & 412ms   & 312ms & 724ms   \\ 
5       & 346ms   & 274ms & 620ms   \\ 
Average & 336.4ms & 243ms & 579.4ms \\ \hline
\end{tabular}
\end{table}
\begin{table}[]
\centering
\caption{Authentication and unlock time}
\label{tab:my-table4}
\begin{tabular}{cccc}
\hline
\textbf{Sl.No} &
  \textbf{\begin{tabular}[c]{@{}c@{}}Time to   create\\    challenge\end{tabular}} &
  \textbf{\begin{tabular}[c]{@{}c@{}}Time to verify   \&\\    authorize\end{tabular}} &
  \textbf{\begin{tabular}[c]{@{}c@{}}Total   time    for\\ processing\end{tabular}} \\ \hline
1       & 212ms   & 319 ms  & 531ms   \\ 
2       & 193ms   & 203ms   & 396ms   \\ 
3       & 224ms   & 324ms   & 548ms   \\ 
4       & 234ms   & 304ms   & 538ms   \\ 
5       & 260ms   & 209ms   & 469ms   \\ 
Average & 224.6ms & 271.8ms & 496.4ms \\ \hline
\end{tabular}
\end{table}
\subsection{Discussion}
Around 45\% of cybercrimes happen due to the lack of awareness regarding the common safeguards among the common people \cite{vijayalakshmi2021impacts}. This is in contrast with the conventional wisdom that an individual is responsible for forestalling their accounts through various practices for example keeping their social media under check or keeping a strong password. This study analyzed the various cyber security threats and after a comprehensive investigation found that FIDO2 specifications are the most reliable user authentication system \cite{klieme2020fidonuous,lyastani2020fido2} and when paired with decentralized e-banking deliver high performance and security. After accounting for other possible solutions, the device attestation methodology excelled because a physical device or a trusted platform is used to store the private keys of the public key cryptosystem, which are used to encrypt important information using RSA, AES128, and AES256 cryptosystems. Since the keys never leave the device, the chances of remote hijack are nullified as well as the user's responsibilities for protecting their account since the biometric information is stored in the device itself. We cannot exclude the possibility of device theft but even then they’ll be unsuccessful to access the account. We believe that the study has provided a novel solution, future work could seek to include additional controls.
\section{Conclusion} \label{topic1:l4}
Decentralised Financial Technology has revolutionized the banking environment and has contributed exponentially to the development of banking technologies. However, on the user end, there lacks several measures for fraud prevention. Many people without the proper knowledge and lack of security abstain from maneuvering the benefits. The purpose of this paper is to propose a sustainable, secure and defensible solution to the greater problem of fraud in decentralized financial technology. The paper reviews the different types of e-frauds committed and tries to provide a wholesome solution that follows the regulations specified by the government and allows the user to securely perform transactions. With the use of cutting-edge technology and FIDO2 specification, it ensures that the user is safe guarded against the various kinds of frauds and attacks, more no one other than the account holder can authorize for access. Further it ensures that the data is available and not used, disclosed, accessed, altered or deleted inappropriately while being stored, retrieved or transmitted. This paper provides an effective, easy-to-use, standardized solution to protect people’s consumer rights and promote the unprivileged and unbanked members of society, large segments of the population, particularly micro, small, and medium-sized companies, to use the modern financial services.


\section*{Acknowledgments}
The authors would like to thank the editors and reviewers. The authors would also like to thank S. V. Kota Reddy, Vice
Chancellor of VIT-AP University, and Jagadish Chandra Mudiganti, Registrar of VIT-AP University for their support. Special thanks to Hari Seetha, Director of the Centre of Excellence, Artificial Intelligence and Robotics (AIR) at VIT-AP University.

\bibliographystyle{unsrt}  
\bibliography{references}  

\end{document}